\documentclass[preprint,eqsecnum,showpacs,aps]{revtex4}
\usepackage[dvips]{graphicx}
\usepackage{amssymb}
\usepackage{amsmath}
\usepackage{amsmath, amsthm, amssymb, mathrsfs}

\begin{document}

\title{Three-Qubit Groverian Measure}

\author{
Eylee Jung, Mi-Ra Hwang, DaeKil Park}

\affiliation{Department of Physics, Kyungnam University, Masan,
631-701, Korea}

\author{Levon Tamaryan}

\affiliation{Physics Department, Yerevan State University,
Yerevan, 375025, Armenia}

\author{Sayatnova Tamaryan}

\affiliation{Theory Department, Yerevan Physics Institute,
Yerevan, 375036, Armenia}

\begin{abstract}
The Groverian measures are analytically computed in various types
of three-qubit states. The final results are also expressed in
terms of local-unitary invariant quantities in each type. This
fact reflects the manifest local-unitary invariance of the
Groverian measure. It is also shown that the analytical
expressions for various types have correct limits to other types.
For some types (type 4 and type 5) we failed to compute the
analytical expression of the Groverian measure in this paper.
However, from the consideration of local-unitary invariants we
have shown that the Groverian measure in type 4 should be
independent of the phase factor $\varphi$, which appear in the
three-qubit state $|\psi \rangle$. This fact with geometric
interpretation on the Groverian measure may enable us to derive
the analytical expressions for general arbitrary three-qubit
states in near future.
\end{abstract}

\pacs{03.67.Mn, 03.65.Ud, 03.67.Bg}

\maketitle

\section{Introduction}

Recently, much attention is paid to quantum
entanglement\cite{keyl02}. It is believed in quantum information
community that entanglement is the physical resource which makes
quantum computer outperforms classical one\cite{vidal03-1}. Thus
in order to exploit fully this physical resource for constructing
and developing quantum algorithms it is important to quantify the
entanglement. The quantity for the quantification is usually
called entanglement measure.

About decade ago the axioms which entanglement measures should
satisfy were studied\cite{vedral97-1}. The most important property
for measure is monotonicity under local operation and classical
communication(LOCC)\cite{vidal98-1}. Following the axioms, many
entanglement measures were constructed such as relative
entropy\cite{plen01-1}, entanglement of distillation\cite{dis96}
and formation\cite{form1,form2,form3,form4}, geometric
measure\cite{shim95,barn01-1,wei03-1,pit}, Schmidt
measure\cite{eisert01} and Groverian measure\cite{biham-1}.
Entanglement measures are used in various branches of quantum
mechanics. Especially, recently, they are used to try to
understand Zamolodchikov's c-theorem\cite{zamo86} more profoundly.
It may be an important application of the quantum information
techniques to understand the effect of renormalization group in
field theories\cite{orus07}.

The purpose of this paper is to compute the Groverian measure for various three-qubit
quantum states.The Groverian measure $G(\psi)$ for three-qubit state $|\psi\rangle$
is defined by
$G(\psi) \equiv \sqrt{1 - P_{max}}$ where
\begin{equation}
\label{pmax1}
P_{max} = \max_{|q_1\rangle,|q_2\rangle,|q_3\rangle} |\langle q_1| \langle q_2|
\langle q_3 | \psi \rangle|^2.
\end{equation}
Thus $P_{max}$ can be interpreted as a maximal overlap between the
given state $|\psi\rangle$ and product states. Groverian measure
is an operational treatment of a geometric measure. Thus, if one
can compute $G(\psi)$, one can also compute the geometric measure
of pure state by $G^2(\psi)$. Sometimes it is more convenient to
re-express Eq.(\ref{pmax1}) in terms of the density matrix $\rho =
|\psi \rangle \langle \psi |$. This can be easily accomplished by
an expression
\begin{equation}
\label{pmax2}
P_{max} = \max_{R^1,R^2,R^3} \mbox{Tr} \left[\rho R^1 \otimes R^2 \otimes R^3
                                                                   \right]
\end{equation}
where $R^i \equiv |q_i \rangle \langle q_i |$ density matrix for the product state.
Eq.(\ref{pmax1}) and Eq.(\ref{pmax2}) manifestly show that $P_{max}$ and $G(\psi)$ are
local-unitary(LU) invariant quantities. Since it is well-known that three-qubit
system has five independent LU-invariants\cite{coff99,acin00,sud00}, {\it say}
$J_i (i = 1, \cdots, 5)$, we would
like to focus on the relation of the Groverian measures to LU-invariants
$J_i$'s in this paper.

This paper is organized as follows. In section II we review simple
case, {\it i.e.} two-qubit system. Using Bloch form of the density
matrix it is shown in this section that two-qubit system has only
one independent LU-invariant quantity, {\it say} $J$. It is also
shown that Groverian measure and $P_{max}$ for arbitrary two-qubit
states can be expressed solely in terms of $J$. In section III we
have discussed how to derive LU-invariants in higher-qubit
systems. In fact, we have derived many LU-invariant quantities
using Bloch form of the density matrix in three-qubit system. It
is shown that all LU-invariants derived can be expressed in terms
of $J_i$'s discussed in Ref.\cite{acin00}. Recently, it was shown
in Ref.\cite{jung07-1} that $P_{max}$ for $n$-qubit state can be
computed from $(n-1)$-qubit reduced mixed state. This theorem was
used in Ref.\cite{tama07-1} and Ref.\cite{tama08-1} to compute
analytically the geometric measures for various three-qubit
states. In this section we have discussed the physical reason why
this theorem is possible from the aspect of LU-invariance. In
section IV we have computed the Groverian measures for various
types of the three-qubit system. The five types we discussed in
this section were originally developed in Ref.\cite{acin00} for
the classification of the three-qubit states. It has been shown
that the Groverian measures for type 1, type 2, and type 3 can be
analytically computed. We have expressed all analytical results in
terms of LU-invariants $J_i$'s. For type 4 and type 5 the
analytical computation seems to be highly nontrivial and may need
separate publications. Thus the analytical calculation for these
types is not presented in this paper. The results of this section
are summarized in Table I. In section V we have discussed the
modified W-like state, which has three-independent real
parameters. In fact, this state cannot be categorized in the five
types discussed in section IV. The analytic expressions of the
Groverian measure for this state was computed recently in
Ref.\cite{tama08-1}. It was shown that the measure has three
different expressions depending on the domains of the parameter
space. It turned out that each expression has its own geometrical
meaning. In this section we have re-expressed all expressions of
the Groverian measure in terms of LU-invariants. In section VI
brief conclusion is given.

\section{Two Qubit: Simple Case}

In this section we consider $P_{max}$ for the two-qubit system.
The Groverian measure for two-qubit system is already
well-known\cite{shim04}. However, we revisit this issue here to
explore how the measure is expressed in terms of the LU-invariant
quantities. The Schmidt decomposition\cite{schmidt1907} makes the
most general expression of the two-qubit state vector to be simple
form
\begin{equation}
\label{s1}
|\psi \rangle = \lambda_0 |00\rangle + \lambda_1 |11 \rangle
\end{equation}
with $\lambda_0, \lambda_1 \geq 0$ and $\lambda_0^2 + \lambda_1^2
= 1$. The density matrix for $|\psi \rangle$ can be expressed in
the Bloch form as following:
\begin{equation}
\label{s2} \rho = |\psi \rangle \langle \psi | = \frac{1}{4}
\left[\openone \otimes \openone + v_{1 \alpha} \sigma_{\alpha}
\otimes \openone + v_{2 \alpha} \openone \otimes \sigma_{\alpha} +
g_{\alpha \beta} \sigma_{\alpha} \otimes \sigma_{\beta} \right],
\end{equation}
where
\begin{eqnarray}
\label{s3}
\vec{v}_1 = \vec{v}_2 = \left(   \begin{array}{c}
                                      0                \\
                                      0                \\
                                  \lambda_0^2 - \lambda_1^2
                                  \end{array}
                                                         \right),
\hspace{1.0cm}
g_{\alpha \beta} = \left(        \begin{array}{ccc}
                          2 \lambda_0 \lambda_1 &    0    &    0    \\
                               0    &   -2 \lambda_0 \lambda_1   &   0    \\
                               0    &    0    &    1
                                  \end{array}
                                                         \right).
\end{eqnarray}

In order to discuss the LU transformation we consider first the
quantity $U \sigma_{\alpha} U^{\dagger}$ where $U$ is $2 \times 2$ unitary
matrix. With direct calculation one can prove easily
\begin{equation}
\label{lu1} U \sigma_{\alpha} U^{\dagger} = {\cal O}_{\alpha
\beta} \sigma_{\beta},
\end{equation}
where the explicit expression of ${\cal O}_{\alpha \beta}$ is given in
appendix A. Since ${\cal O}_{\alpha \beta}$ is a real matrix satisfying
${\cal O} {\cal O}^T = {\cal O}^T {\cal O} = \openone$, it is an element of the
rotation group O(3). Therefore, Eq.(\ref{lu1}) implies that the LU-invariants
in the density
matrix (\ref{s2}) are $|\vec{v}_1|$, $|\vec{v}_2|$,
$\mbox{Tr}[g g^T]$ etc.

All LU-invariant quantities can be written in terms of one
quantity, {\it say} $J \equiv \lambda_0^2 \lambda_1^2$. In fact,
$J$ can be expressed in terms of two-qubit concurrence\cite{form3}
${\cal C}$ by ${\cal C}^2 / 4$. Then it is easy to show
\begin{eqnarray}
\label{s4} & &|\vec{v}_1|^2 = |\vec{v}_2|^2 = 1 - 4 J,    \\
\nonumber & & g_{\alpha \beta} g_{\alpha \beta} = 1 + 8 J.
\end{eqnarray}

It is well-known that $P_{max}$ is simply square of larger Schmidt
number in two-qubit case

\begin{equation}
\label{s7} P_{max} = \mbox{max} \left( \lambda_0^2, \lambda_1^2
\right).
\end{equation}

It can be re-expressed in terms of reduced density operators

\begin{equation}
\label{s5} P_{max} = \frac{1}{2} \left[1 + \sqrt{1 - 4 \mbox{det}
\rho^A} \right],
\end{equation}

\noindent where $\rho^A = \mbox{Tr}_B \rho = (1 + v_{1 \alpha}
\sigma_{\alpha})/2$. Since $P_{max}$ is invariant under
LU-transformation, it should be expressed in terms of LU-invariant
quantities. In fact, $P_{max}$ in Eq.(\ref{s5}) can be re-written
as
\begin{equation}
\label{s6}
P_{max} = \frac{1}{2} \left[1 + \sqrt{1 - 4 J}\right].
\end{equation}
Eq.(\ref{s6}) implies that $P_{max}$ is manifestly LU-invariant.

\section{Local Unitary Invariants}
The Bloch representation of the $3$-qubit density matrix can be written in the
form
\begin{eqnarray}
\label{density1}
\rho = \frac{1}{8}
& &\Bigg[\openone \otimes \openone \otimes \openone + v_{1 \alpha}
\sigma_{\alpha}\otimes \openone \otimes \openone + v_{2 \alpha}
\openone \otimes \sigma_{\alpha} \otimes \openone + v_{3 \alpha}
\openone \otimes \openone \otimes \sigma_{\alpha}
                                                              \\   \nonumber
& &+ h^{(1)}_{\alpha \beta} \openone \otimes
\sigma_{\alpha}\otimes \sigma_{\beta} + h^{(2)}_{\alpha \beta}
\sigma_{\alpha}\otimes \openone \otimes \sigma_{\beta} +
h^{(3)}_{\alpha \beta} \sigma_{\alpha}\otimes \sigma_{\beta}
\otimes \openone + g_{\alpha \beta \gamma} \sigma_{\alpha}\otimes
\sigma_{\beta} \otimes \sigma_{\gamma} \Bigg], \vspace{1cm}
\end{eqnarray}
where $\sigma_{\alpha}$ is Pauli matrix.
According to Eq.(\ref{lu1}) and  appendix A it is easy to show that
the LU-invariants in the density
matrix (\ref{density1}) are $|\vec{v}_1|$, $|\vec{v}_2|$, $|\vec{v}_3|$,
$\mbox{Tr}[h^{(1)} h^{(1) T}]$, $\mbox{Tr}[h^{(2)} h^{(2) T}]$,
$\mbox{Tr}[h^{(3)} h^{(3) T}]$, $g_{\alpha \beta \gamma} g_{\alpha \beta \gamma}$
etc.

Few years ago Ac\'in et al\cite{acin00} represented the
three-qubit arbitrary states in a simple form using a generalized
Schmidt decomposition\cite{schmidt1907} as following:
\begin{equation}
\label{state1}
|\psi\rangle = \lambda_0 |000\rangle + \lambda_1 e^{i \varphi} |100\rangle
+ \lambda_2 |101\rangle + \lambda_3 |110\rangle + \lambda_4 |111\rangle
\end{equation}
with $\lambda_i \geq 0$, $0 \leq \varphi \leq \pi$, and $\sum_i
\lambda_i^2 = 1$. The five algebraically independent
polynomial LU-invariants were also constructed in
Ref.\cite{acin00}:
\begin{eqnarray}
\label{lu2} & &J_1 = \lambda_1^2 \lambda_4^2 + \lambda_2^2
\lambda_3^2 - 2 \lambda_1 \lambda_2 \lambda_3 \lambda_4 \cos
\varphi,
                                                      \\  \nonumber
& &J_2 = \lambda_0^2 \lambda_2^2, \hspace{1.0cm}
   J_3 = \lambda_0^2 \lambda_3^2, \hspace{1.0cm}
   J_4 = \lambda_0^2 \lambda_4^2,
                                                      \\  \nonumber
& &J_5 = \lambda_0^2 (J_1 + \lambda_2^2 \lambda_3^2 - \lambda_1^2 \lambda_4^2).
\end{eqnarray}

In order to determine how many states have the same
values of the invariants $J_1, J_2, ...J_5$, and therefore how
many further discrete-valued invariants are needed to specify
uniquely a pure state of three qubits up to local transformations,
one would need to find the number of different sets of parameters
$\varphi$ and $\lambda_i(i=0,1,...4)$, yielding the same
invariants. Once $\lambda_0$ is found, other parameters are
determined uniquely and therefore we derive an equation defining
$\lambda_0$ in terms of polynomial invariants.

\begin{equation}\label{added}
(J_1+J_4)\lambda_0^4-(J_5+J_4)\lambda_0^2+J_2J_3+J_2J_4+J_3J_4+J_4^2=0.
\end{equation}

This equation has at most two positive roots and
consequently an additional discrete-valued invariant is required
to specify uniquely a pure three qubit state. Generally 18
LU-invariants, nine of which may be taken to have only discrete
values, are needed to determine a mixed 2-qubit state
\cite{mixed}.

If one represents the density matrix $|\psi \rangle \langle \psi
|$ as a Bloch form like Eq.(\ref{density1}), it is possible to
construct $v_{1 \alpha}$, $v_{2 \alpha}$, $v_{3 \alpha}$,
$h^{(1)}_{\alpha \beta}$, $h^{(2)}_{\alpha \beta}$,
$h^{(3)}_{\alpha \beta}$, and $g_{\alpha \beta \gamma}$
explicitly, which are summarized in appendix B. Using these
explicit expressions one can show directly that all polynomial
LU-invariant quantities of pure states are expressed in terms of
$J_i$ as following:
\begin{eqnarray}
\label{lu3}
& &|\vec{v}_1|^2 = 1 - 4 (J_2 + J_3 + J_4), \hspace{1.0cm}
|\vec{v}_2|^2 = 1 - 4 (J_1 + J_3 + J_4)
                                              \\   \nonumber
& &|\vec{v}_3|^2 = 1 - 4 (J_1 + J_2 + J_4), \hspace{1.0cm}
\mbox{Tr}[h^{(1)} h^{(1) T}] = 1 + 4 (2 J_1 - J_2 - J_3)
                                               \\   \nonumber
& &\mbox{Tr}[h^{(2)} h^{(2) T}] = 1 - 4 ( J_1 - 2 J_2 + J_3), \hspace{.5cm}
\mbox{Tr}[h^{(3)} h^{(3) T}] = 1 - 4 (J_1 + J_2 - 2 J_3)
                                                \\  \nonumber
& &g_{\alpha \beta \gamma} g_{\alpha \beta \gamma}
= 1 + 4(2 J_1 + 2 J_2 + 2 J_3 + 3 J_4)
                                               \\   \nonumber
& &h^{(3)}_{\alpha \beta} v^{(1)}_{\alpha} v^{(2)}_{\beta} =
1 - 4 (J_1 + J_2 + J_3 + J_4 - J_5).
\end{eqnarray}

Recently, Ref.\cite{jung07-1} has shown that $P_{max}$ for
$n$-qubit pure state can be computed from $(n-1)$-qubit reduced
mixed state. This is followed from a fact
\begin{equation}
\label{theorem1}
\max_{R^1,R^2\cdots R^n}
\mbox{Tr}\left[\rho R^1 \otimes R^2 \otimes \cdots \otimes R^n \right]
= \max_{R^1, R^2 \cdots R^{n-1}}
\mbox{Tr}\left[\rho R^1 \otimes R^2 \otimes \cdots \otimes R^{n-1} \otimes \openone
                                                                      \right]
\end{equation}
which is Theorem I of Ref.\cite{jung07-1}. Here, we would like to discuss the
physical meaning of Eq.(\ref{theorem1}) from the aspect of LU-invariance.
Eq.(\ref{theorem1}) in $3$-qubit system reduces to
\begin{equation}
\label{pmax21}
P_{max} = \max_{R^1,R^2} \mbox{Tr} \left[\rho^{AB} R^1 \otimes R^2 \right]
\end{equation}
where $\rho^{AB} = \mbox{Tr}_C \rho$. From Eq.(\ref{density1}) $\rho^{AB}$
simply reduces to
\begin{equation}
\label{density21}
\rho =
\frac{1}{4} \left[\openone \otimes \openone + v_{1 \alpha}
\sigma_{\alpha} \otimes \openone + v_{2 \alpha} \openone \otimes \sigma_{\alpha}
+ h^{(3)}_{\alpha \beta} \sigma_{\alpha} \otimes \sigma_{\beta} \right]
\end{equation}
where $ v_{1 \alpha}$, $v_{2 \alpha}$ and $ h^{(3)}_{\alpha \beta}$ are explicitly
given in appendix B. Of course, the LU-invariant quantities of $\rho^{AB}$ are
$|\vec{v}_1|$, $|\vec{v}_2|$, $\mbox{Tr}[h^{(3)} h^{(3) T}]$,
$h^{(3)}_{\alpha \beta} v_{1 \alpha} v_{2 \beta}$ etc, all of which, of course,
can be re-expressed in terms of $J_1$, $J_2$, $J_3$, $J_4$ and $J_5$.
It is worthwhile noting that we need all $J_i$'s to express the LU-invariant
quantities of $\rho^{AB}$. This means that the reduced state $\rho^{AB}$ does
have full information on the LU-invariance of the original pure state $\rho$.

Indeed, any reduced state resulting from a partial trace over a
single qubit uniquely determines any entanglement measure of
original system, given that the initial state is pure. Consider an
$(n-1)$-qubit reduced density matrix that can be purified by a
single qubit reference system. Let $|\psi^\prime\rangle$ be any
joint pure state. All other purifications can be obtained from the
state $|\psi^\prime\rangle$ by LU-transformations
$U\otimes\openone^{\otimes(n-1)}$, where $U$ is a local unitary
matrix acting on single qubit. Since any entanglement measure must
be invariant under LU-transformations, it must be same for all
purifications independently of $U$. Hence the reduced density
matrix determines any entanglement measure on the initial pure
state. That is why we can compute $P_{max}$ of $n$-qubit pure
state from the $(n-1)$-qubit reduced mixed state.

Generally, the information on the LU-invariance
of the original $n$-qubit state is partly lost if we take partial trace twice.
In order to show this explicitly let us consider
$\rho^A \equiv \mbox{Tr}_B \rho^{AB}$ and
$\rho^B \equiv \mbox{Tr}_A \rho^{AB}$:
\begin{eqnarray}
\label{density22}
\rho^A&=&\frac{1}{2} \left[ \openone + v_{1 \alpha} \sigma_{\alpha} \right]
                                                              \\   \nonumber
\rho^B&=&\frac{1}{2} \left[ \openone + v_{2 \alpha} \sigma_{\alpha} \right].
\end{eqnarray}
Eq.(\ref{lu1}) and appendix A imply that their LU-invariant quantities are
only $|\vec{v}_1|$ and $|\vec{v}_2|$ respectively. Thus, we do not need
$J_5$ to express the LU-invariant quantities of $\rho^A$ and $\rho^B$.
This fact indicates that the mixed states $\rho^A$ and $\rho^B$ partly
loose the information of
the LU-invariance of the original pure state $\rho$. This is why
$(n-2)$-qubit reduced state cannot be used to compute $P_{max}$ of
$n$-qubit pure state.

\section{Calculation of $P_{max}$}

\subsection{General Feature}
If we insert the Bloch representation
\begin{equation}
\label{bloch41}
R^1 = \frac{\openone + \vec{s}_1 \cdot \vec{\sigma}}{2}
\hspace{1.0cm}
R^2 = \frac{\openone + \vec{s}_2 \cdot \vec{\sigma}}{2}
\end{equation}
with $|\vec{s}_1| = |\vec{s}_2| = 1$ into Eq.(\ref{pmax21}), $P_{max}$ for
$3$-qubit state becomes
\begin{equation}
\label{pmax41}
P_{max} = \frac{1}{4} \max_{|\vec{s}_1| = |\vec{s}_2| = 1}
\left[1 + \vec{r}_1 \cdot \vec{s}_1 + \vec{r}_2 \cdot \vec{s}_2 +
     g_{i j} s_{1 i} s_{2 j} \right]
\end{equation}
where
\begin{eqnarray}
\label{def41}
& &\vec{r}_1 = \mbox{Tr} \left[\rho^A \vec{\sigma} \right]
                                                         \\   \nonumber
& &\vec{r}_2 = \mbox{Tr} \left[\rho^B \vec{\sigma} \right]
                                                         \\   \nonumber
& &g_{ij} = \mbox{Tr} \left[\rho^{AB} \sigma_i \otimes \sigma_j \right].
\end{eqnarray}
Since in Eq.(\ref{pmax41}) $P_{max}$ is maximization with constraint
 $|\vec{s}_1| = |\vec{s}_2| = 1$, we should use the Lagrange multiplier method,
which yields a pair of equations
\begin{eqnarray}
\label{algebraic1} & &\vec{r}_1 + g \vec{s}_2 = \Lambda_1
\vec{s}_1          \\   \nonumber & &\vec{r}_2 + g^T \vec{s}_1 =
\Lambda_2 \vec{s}_2,
\end{eqnarray}
where the symbol $g$ represents the matrix $g_{ij}$ in Eq.(\ref{def41}). Thus we
should solve $\vec{s}_1$, $\vec{s}_2$, $\Lambda_1$ and $\Lambda_2$
by eq.(\ref{algebraic1}) and the constraint $|\vec{s}_1| =
|\vec{s}_2| = 1$. Although it is highly nontrivial to solve
Eq.(\ref{algebraic1}), sometimes it is not difficult if the given
$3$-qubit state $|\psi\rangle$ has rich symmetries. Now, we would
like to compute $P_{max}$ for various types of $3$-qubit system.

\subsection{Type 1 (Product States): $J_1 = J_2 = J_3 = J_4 = J_5 = 0$}
In order for all $J_i$'s to be zero we have two cases $\lambda_0 = J_1 = 0$ or
$\lambda_2 = \lambda_3 = \lambda_4 = 0$.
\subsubsection{$\lambda_0 = J_1 = 0$}
If $\lambda_0 = 0$, $|\psi \rangle$ in Eq.(\ref{state1}) becomes
$|\psi \rangle = |1\rangle \otimes |BC \rangle$
where
\begin{equation}
\label{bc41}
|BC\rangle = \lambda_1 e^{i \varphi} |00\rangle + \lambda_2 |01\rangle
+ \lambda_3 |10\rangle + \lambda_4 |11\rangle.
\end{equation}
Thus $P_{max}$ for $|\psi\rangle$ equals to that for $|BC\rangle$. Since
$|BC\rangle$ is two-qubit state, one can easily compute $P_{max}$ using
Eq.(\ref{s5}), which is
\begin{equation}
\label{pmax42}
P_{max} = \frac{1}{2} \left[1 + \sqrt{1 - 4 \mbox{det}
           \left(\mbox{Tr}_B |BC\rangle \langle BC| \right)} \right]
        = \frac{1}{2} \left[1 + \sqrt{1 - 4 J_1}\right].
\end{equation}
If, therefore, $\lambda_0 = J_1 = 0$, we have $P_{max} = 1$, which gives a
vanishing Groverian measure.
\subsubsection{$\lambda_2 = \lambda_3 = \lambda_4 = 0$}
In this case $|\psi \rangle$ in Eq.(\ref{state1}) becomes
\begin{equation}
\label{psi41}
|\psi \rangle = \left(\lambda_0 |0\rangle + \lambda_1 e^{i \varphi} |1\rangle\right)
\otimes |0\rangle \otimes |0\rangle.
\end{equation}
Since $|\psi \rangle$ is completely product state, $P_{max}$ becomes one.
\subsection{Type2a (biseparable states)}
In this type we have following three cases.
\subsubsection{$J_1 \neq 0$ and $J_2=J_3=J_4=J_5=0$}
In this case we have $\lambda_0 = 0$. Thus $P_{max}$ for this case is
exactly same with Eq.(\ref{pmax42}).
\subsubsection{$J_2 \neq 0$ and $J_1=J_3=J_4=J_5=0$}
In this case we have $\lambda_2 = \lambda_4 = 0$. Thus $P_{max}$ for $|\psi\rangle$
equals to that for $|AC\rangle$, where
\begin{equation}
\label{ac41}
|AC\rangle = \lambda_0 |00\rangle + \lambda_1 e^{i \varphi} |10\rangle +
\lambda_2 |11\rangle.
\end{equation}
Using Eq.(\ref{s5}), therefore, one can easily compute $P_{max}$, which is
\begin{equation}
\label{pmax43}
P_{max} = \frac{1}{2} \left[1 + \sqrt{1 - 4 J_2}\right].
\end{equation}
\subsubsection{$J_3 \neq 0$ and $J_1=J_2=J_4=J_5=0$}
In this case $P_{max}$ for $|\psi\rangle$
equals to that for $|AB\rangle$, where
\begin{equation}
\label{ab41}
|AB\rangle = \lambda_0 |00\rangle + \lambda_1 e^{i \varphi} |10\rangle +
\lambda_3 |11\rangle.
\end{equation}
Thus $P_{max}$ for $|\psi\rangle$ is
\begin{equation}
\label{pmax44}
P_{max} = \frac{1}{2} \left[1 + \sqrt{1 - 4 J_3}\right].
\end{equation}

\subsection{Type2b (generalized GHZ states): $J_4 \neq 0$, $J_1=J_2=J_3=J_5=0$}
In this case we have $\lambda_1 = \lambda_2 = \lambda_3 = 0$ and $|\psi\rangle$
becomes
\begin{equation}
\label{ghz41}
|\psi \rangle = \lambda_0 |000\rangle + \lambda_4 |111\rangle
\end{equation}
with $\lambda_0^2 + \lambda_4^2 = 1$. Then it is easy to show
\begin{eqnarray}
\label{def42}
& &\vec{r}_1 = \mbox{Tr} \left[\rho^A \vec{\sigma} \right] =
(0, 0, \lambda_0^2 - \lambda_4^2)
                                                        \\   \nonumber
& &\vec{r}_2 = \mbox{Tr} \left[\rho^B \vec{\sigma} \right] =
(0, 0, \lambda_0^2 - \lambda_4^2)
                                                        \\    \nonumber
& &g_{ij} = \mbox{Tr} \left[\rho^{AB} \sigma_i \otimes \sigma_j \right]
= \left(            \begin{array}{ccc}
                    0  &  0  &  0    \\
                    0  &  0  &  0    \\
                    0  &  0  &  1
                    \end{array}            \right).
\end{eqnarray}
Thus $P_{max}$ reduces to
\begin{equation}
\label{pmax45}
P_{max} = \frac{1}{4} \max_{|\vec{s}_1| = |\vec{s}_2| = 1}
\left[1 + (\lambda_0^2 - \lambda_4^2) (s_{1 z} + s_{2 z}) + s_{1 z} s_{2 z} \right].
\end{equation}
Since Eq.(\ref{pmax45}) is simple, we do not need to solve Eq.(\ref{algebraic1}) for
the maximization. If $\lambda_0 > \lambda_4$, the maximization can be achieved by simply
choosing $\vec{s}_1 = \vec{s}_2 = (0, 0, 1)$. If $\lambda_0 < \lambda_4$, we
choose $\vec{s}_1 = \vec{s}_2 = (0, 0, -1)$. Thus we have
\begin{equation}
\label{pmax46}
P_{max} = \mbox{max} (\lambda_0^2, \lambda_4^2).
\end{equation}

In order to express $P_{max}$ in Eq.(\ref{pmax46}) in terms of LU-invariants we
follow the following procedure. First we note
\begin{equation}
\label{pmax47}
P_{max} = \frac{1}{2} \left[ (\lambda_0^2 + \lambda_4^2) + |\lambda_0^2 - \lambda_4^2|
                                                                        \right].
\end{equation}
Since $|\lambda_0^2 - \lambda_4^2| = \sqrt{(\lambda_0^2 + \lambda_4^2)^2 -
         4 \lambda_0^2 \lambda_4^2} = \sqrt{1 - 4 J_4}$, we get finally
\begin{equation}
\label{pmax48}
P_{max} = \frac{1}{2} \left[1 + \sqrt{1 - 4 J_4}\right].
\end{equation}

\subsection{Type3a (tri-Bell states)}
In this case we have $\lambda_1 = \lambda_4 = 0$ and $|\psi\rangle$ becomes
\begin{equation}
\label{w41}
|\psi \rangle = \lambda_0 |000\rangle + \lambda_2 |101\rangle +
\lambda_3 |110\rangle
\end{equation}
with $\lambda_0^2 + \lambda_2^2 + \lambda_3^2 = 1$. If we take LU-transformation
$\sigma_x$ in the first-qubit, $|\psi\rangle$ is changed into
$|\psi'\rangle$ which is usual W-type state\cite{dur00} as follows:
\begin{equation}
\label{w42}
|\psi'\rangle = \lambda_0 |100\rangle + \lambda_3 |010\rangle +
\lambda_2 |001\rangle.
\end{equation}
The LU-invariants in this type are
\begin{eqnarray}
\label{lu41}
& & J_1 = \lambda_2^2 \lambda_3^2 \hspace{1.0cm}
    J_2 = \lambda_0^2 \lambda_2^2
                                        \\   \nonumber
& & J_3 =  \lambda_0^2 \lambda_3^2 \hspace{1.0cm}
    J_5 = 2 \lambda_0^2 \lambda_2^2 \lambda_3^2.
\end{eqnarray}
Then it is easy to derive a relation
\begin{equation}
\label{rela41}
J_1 J_2 + J_1 J_3 + J_2 J_3 = \sqrt{J_1 J_2 J_3} = \frac{1}{2} J_5.
\end{equation}

Recently, $P_{max}$ for $|\psi'\rangle$ is computed analytically in Ref.\cite{tama07-1}
by solving the Lagrange multiplier equations (\ref{algebraic1}) explicitly.
In order to express $P_{max}$ explicitly we
first define
\begin{eqnarray}
\label{def43}
r_1&=& \lambda_3^2 + \lambda_2^2 - \lambda_0^2   \\   \nonumber
r_2&=& \lambda_0^2 + \lambda_2^2 - \lambda_3^2   \\   \nonumber
r_3&=& \lambda_0^2 + \lambda_3^2 - \lambda_2^2   \\   \nonumber
\omega&=&2 \lambda_0 \lambda_3.
\end{eqnarray}
Also we define
\begin{eqnarray}
\label{def44}
a&=&\mbox{max} (\lambda_0, \lambda_2, \lambda_3)    \\    \nonumber
b&=&\mbox{mid} (\lambda_0, \lambda_2, \lambda_3)    \\    \nonumber
c&=&\mbox{min} (\lambda_0, \lambda_2, \lambda_3).
\end{eqnarray}
Then $P_{max}$ is expressed differently in two different regions as follows.
If $a^2 \geq b^2 + c^2$, $P_{max}$ becomes
\begin{equation}
\label{pmax49}
P_{max}^{>} = a^2 = \mbox{max}(\lambda_0^2, \lambda_2^2, \lambda_3^2).
\end{equation}
In order to express $P_{max}$ in terms of LU-invariants we express Eq.(\ref{pmax49})
differently as
\begin{equation}
\label{pmax50}
P_{max}^{>} = \frac{1}{4}
\left[(\lambda_0^2 + \lambda_3^2 + \lambda_2^2)
       + |\lambda_0^2 + \lambda_3^2 - \lambda_2^2|
       + |\lambda_0^2 - \lambda_3^2 + \lambda_2^2|
       + |\lambda_0^2 - \lambda_3^2 - \lambda_2^2|  \right].
\end{equation}
Using equalities
\begin{eqnarray}
\label{equal41}
& & |\lambda_0^2 + \lambda_3^2 - \lambda_2^2| =
\sqrt{1 - 4 \lambda_0^2 \lambda_2^2 - 4 \lambda_2^2 \lambda_3^2}
= \sqrt{1 - 4(J_1 + J_2)}
                                                      \\   \nonumber
& & |\lambda_0^2 - \lambda_3^2 + \lambda_2^2| =
\sqrt{1 - 4 \lambda_0^2 \lambda_3^2 - 4 \lambda_2^2 \lambda_3^2}
= \sqrt{1 - 4(J_1 + J_3)}
                                                      \\   \nonumber
& & |\lambda_0^2 - \lambda_3^2 - \lambda_2^2| =
\sqrt{1 - 4 \lambda_0^2 \lambda_2^2 - 4 \lambda_0^2 \lambda_3^2}
= \sqrt{1 - 4(J_2 + J_3)},
\end{eqnarray}
we can express $P_{max}$ in Eq.(\ref{pmax49}) as follows:
\begin{equation}
\label{pmax51}
P_{max}^{>} = \frac{1}{4}
\left[1 + \sqrt{1 - 4(J_1 + J_2)} + \sqrt{1 - 4(J_1 + J_3)} +  \sqrt{1 - 4(J_2 + J_3)}
                                                   \right].
\end{equation}

If $a^2 \leq b^2 + c^2$, $P_{max}$ becomes
\begin{equation}
\label{pmax52}
P_{max}^{<} = \frac{1}{4}
\left[1 + \frac{\omega \sqrt{(\omega^2 + r_1^2 - r_3^2)(\omega^2 + r_2^2 - r_3^2)}
                - r_1 r_2 r_3}
               {\omega^2 - r_3^2}          \right].
\end{equation}
It was shown in Ref.\cite{tama07-1} that $P_{max} = 4 R^2$, where
$R$ is a circumradius of the triangle $\lambda_0$, $\lambda_2$ and
$\lambda_3$. When $a^2 \leq b^2 + c^2$, one can show easily $r_1 =
\sqrt{1 - 4(J_2 + J_3)}$, $r_2 = \sqrt{1 - 4(J_1 + J_3)}$, $r_3 =
\sqrt{1 - 4(J_1 + J_2)}$, and $\omega = 2 \sqrt{J_3}$. Using
$\omega^2 - r_3^2 - r_1 r_2 r_3 = 8 \lambda_0^2
\lambda_2^2\lambda_3^2$, One can show easily that $P_{max}$ in
Eq.(\ref{pmax52}) in terms of LU-invariants becomes
\begin{equation}
\label{pmax53}
P_{max}^{<} = \frac{4 \sqrt{J_1 J_2 J_3}}{4 (J_1 + J_2 + J_3) - 1}.
\end{equation}

Let us consider $\lambda_0 = 0$ limit in this type. Then we have $J_2 = J_3 = 0$.
Thus $P_{max}^{>}$ reduces to $(1/2) (1 + \sqrt{1 - 4 J_1})$ which exactly coincides
with Eq.(\ref{pmax42}). By same way one can prove that Eq.(\ref{pmax51}) has correct
limits to various other types.

\subsection{Type3b (extended GHZ states)}
This type consists of $3$ types, {\it i.e.} $\lambda_1 = \lambda_2=0$,
$\lambda_1=\lambda_3=0$ and $\lambda_2=\lambda_3=0$.

\subsubsection{$\lambda_1 = \lambda_2 = 0$}
In this case the state (\ref{state1}) becomes
\begin{equation}
\label{eghz41}
|\psi \rangle = \lambda_0 |000\rangle + \lambda_3 |110\rangle + \lambda_4 |111\rangle
\end{equation}
with $\lambda_0^2 + \lambda_3^2 + \lambda_4^2 = 1$. The non-vanishing LU-invariants
are
\begin{equation}
\label{eghzlu1}
J_3 = \lambda_0^2 \lambda_3^2, \hspace{1.0cm} J_4 = \lambda_0^2 \lambda_4^2.
\end{equation}
Note that $J_3 + J_4$ is expressed in terms of solely $\lambda_0$ as
\begin{equation}
\label{eghzlu2}
J_3 + J_4 = \lambda_0^2 (1 - \lambda_0^2).
\end{equation}

Eq.(\ref{eghz41}) can be re-written as
\begin{equation}
\label{eghz42}
|\psi \rangle = \lambda_0 |00q_1 \rangle + \sqrt{1 - \lambda_0^2} |11q_2 \rangle
\end{equation}
where $|q_1\rangle = |0\rangle$ and
$|q_2\rangle = (1 / \sqrt{1 - \lambda_0^2}) (\lambda_3 |0\rangle + \lambda_4|1\rangle)$
are normalized one qubit states. Thus, from Ref.\cite{tama07-1}, $P_{max}$ for
$|\psi \rangle$ is
\begin{equation}
\label{eghzpmax1}
P_{max} = \mbox{max} \left(\lambda_0^2, 1 - \lambda_0^2 \right)
= \frac{1}{2} \left[1 + \sqrt{(1 - 2 \lambda_0^2)^2} \right].
\end{equation}
With an aid of Eq.(\ref{eghzlu2}) $P_{max}$ in Eq.(\ref{eghzpmax1}) can be easily
expressed in terms of LU-invariants as following:
\begin{equation}
\label{eghzpmax2}
P_{max} = \frac{1}{2} \left[1 + \sqrt{1 - 4 (J_3 + J_4)} \right].
\end{equation}
If we take $\lambda_3 = 0$ limit in this type, we have $J_3 = 0$, which makes
Eq.(\ref{eghzpmax2}) to be $(1/2) (1 + \sqrt{1 - 4 J_4})$. This exactly coincides with
Eq.(\ref{pmax48}).

\subsubsection{$\lambda_1 = \lambda_3 = 0$}
In this case $|\psi\rangle$ and LU-invariants are
\begin{equation}
\label{eghz43}
|\psi\rangle = \lambda_0 |0 q_1 0\rangle + \sqrt{1 - \lambda_0^2}
|1 q_2 1\rangle
\end{equation}
and
\begin{equation}
\label{eghzlu3}
J_2 = \lambda_0^2 \lambda_2^2, \hspace{1.0cm} J_4 = \lambda_0^2 \lambda_4^2
\end{equation}
where $|q_1\rangle = |0\rangle$,
$|q_2\rangle = (1 /\sqrt{1 - \lambda_0^2})(\lambda_2 |0\rangle + \lambda_4 |1\rangle)$,
and $\lambda_0^2 + \lambda_2^2 + \lambda_4^2=1$. The same method used in the previous
subsection easily yields
\begin{equation}
\label{eghzpmax3}
P_{max} = \frac{1}{2} \left[1 + \sqrt{1 - 4 (J_2 + J_4)} \right].
\end{equation}
One can show that Eq.(\ref{eghzpmax3}) has correct limits to other types.

\subsubsection{$\lambda_2 = \lambda_3 = 0$}
In this case $|\psi\rangle$ and LU-invariants are
\begin{equation}
\label{eghz44}
|\psi\rangle = \sqrt{1 - \lambda_4^2} |q_1 0 0\rangle +  \lambda_4
|q_2 1 1\rangle
\end{equation}
and
\begin{equation}
\label{eghzlu4}
J_1 = \lambda_1^2 \lambda_4^2, \hspace{1.0cm} J_4 = \lambda_0^2 \lambda_4^2
\end{equation}
where $|q_1\rangle = (1 / \sqrt{1 - \lambda_4^2}) (\lambda_0 |0\rangle +
\lambda_1 e^{i \varphi} |1\rangle)$, $|q_2\rangle = |1\rangle$, and
$\lambda_0^2 + \lambda_1^2 + \lambda_4^2=1$. It is easy to show
\begin{equation}
\label{eghzpmax4}
P_{max} = \frac{1}{2} \left[1 + \sqrt{1 - 4 (J_1 + J_4)} \right].
\end{equation}
One can show that Eq.(\ref{eghzpmax4}) has correct limits to other types.

\subsection{Type4a ($\lambda_4 = 0$)}
In this case the state vector $|\psi \rangle$ in Eq.(\ref{state1}) reduces to
\begin{equation}
\label{4a1}
|\psi\rangle = \lambda_0 |000\rangle + \lambda_1 e^{i \varphi} |100\rangle
+ \lambda_2 |101 \rangle + \lambda_3 |110\rangle
\end{equation}
with $\lambda_0^2 + \lambda_1^2 + \lambda_2^2 + \lambda_3^2 = 1$.
The non-vanishing LU-invariants are
\begin{eqnarray}
\label{4alu1}
& &J_1 = \lambda_2^2 \lambda_3^2 \hspace{1.0cm} J_2 = \lambda_0^2 \lambda_2^2
                                                              \\   \nonumber
& &J_3 = \lambda_0^2 \lambda_3^2 \hspace{1.0cm}
J_5 = 2 \lambda_0^2 \lambda_2^2 \lambda_3^2.
\end{eqnarray}
From Eq.(\ref{4alu1}) it is easy to show
\begin{equation}
\label{4alu2}
\sqrt{J_1 J_2 J_3} = \frac{1}{2} J_5.
\end{equation}
The remarkable fact deduced from Eq.(\ref{4alu1}) is that the non-vanishing LU-invariants
are independent of the phase factor $\varphi$. This indicates that the Groverian measure
for Eq.(\ref{4a1}) is also independent of $\varphi$

In order to compute $P_{max}$ analytically in this type, we should solve the Lagrange
multiplier equations (\ref{algebraic1}) with
\begin{eqnarray}
\label{rrg1}
& &\vec{r}_1 = \mbox{Tr} [\rho^A \vec{\sigma}] = (2 \lambda_0 \lambda_1 \cos \varphi,
2 \lambda_0 \lambda_1 \sin \varphi, 2 \lambda_0^2 - 1)
                                                              \\   \nonumber
& &\vec{r}_2 = \mbox{Tr} [\rho^B \vec{\sigma}] = (2 \lambda_1 \lambda_3 \cos \varphi,
 -2\lambda_1 \lambda_3 \sin \varphi, 1 - 2 \lambda_3^2)
                                                              \\   \nonumber
& & g_{ij} = \mbox{Tr}[\rho^{AB} \sigma_i \otimes \sigma_j] =
\left(                    \begin{array}{ccc}
      2 \lambda_0 \lambda_3       &     0     &    2 \lambda_0 \lambda_1 \cos \varphi  \\
         0     &    -2 \lambda_0 \lambda_3   &   2 \lambda_0 \lambda_1 \sin \varphi   \\
       -2 \lambda_1 \lambda_3 \cos \varphi  &  2 \lambda_1 \lambda_3 \sin \varphi  &
        \lambda_0^2 - \lambda_1^2 - \lambda_2^2 + \lambda_3^2
                          \end{array}
                                                                  \right).
\end{eqnarray}
Although we have freedom to choose the phase factor $\varphi$, it
is impossible to find singular values of the matrix $g$, which
makes it formidable task to solve Eq.(\ref{algebraic1}). Based on
Ref.\cite{tama07-1} and Ref.\cite{tama08-1}, furthermore, we can
conjecture that $P_{max}$ for this type may have several different
expressions depending on the domains in parameter space.
Therefore, it may need long calculation to compute $P_{max}$
analytically. We would like to leave this issue for our future
research work and the explicit expressions of $P_{max}$ are not
presented in this paper.

\subsection{Type4b}
This type consists of the $2$ cases, {\it i.e.} $\lambda_2=0$ and $\lambda_3=0$.

\subsubsection{$\lambda_2=0$}
In this case the state vector $|\psi \rangle$ in Eq.(\ref{state1})
reduces to
\begin{equation}
\label{4b1}
|\psi\rangle = \lambda_0 |000\rangle + \lambda_1 e^{i \varphi} |100\rangle
+ \lambda_3 |110 \rangle + \lambda_4 |111\rangle
\end{equation}
with $\lambda_0^2 + \lambda_1^2 + \lambda_3^2 + \lambda_4^2 = 1$.
The LU-invariants are
\begin{equation}
\label{4blu1}
J_1 = \lambda_1^2 \lambda_4^2 \hspace{.5cm}
J_3 = \lambda_0^2 \lambda_3^2 \hspace{.5cm}
J_4 = \lambda_0^2 \lambda_4^2.
\end{equation}
Eq.(\ref{4blu1}) implies that the Groverian measure for
Eq.(\ref{4b1}) is independent of the phase factor $\varphi$ like
type 4a. This fact may drastically reduce the calculation
procedure for solving the Lagrange multiplier equation
(\ref{algebraic1}). In spite of this fact, however, solving
Eq.(\ref{algebraic1}) is highly non-trivial as we commented in the
previous type. The explicit expressions of the Groverian measure
are not presented in this paper and we hope to present them
elsewhere in the near future.

\subsubsection{$\lambda_3=0$}
In this case the state vector $|\psi \rangle$ in Eq.(\ref{state1})
reduces to
\begin{equation}
\label{4b2}
|\psi\rangle = \lambda_0 |000\rangle + \lambda_1 e^{i \varphi} |100\rangle
+ \lambda_2 |101 \rangle + \lambda_4 |111\rangle
\end{equation}
with $\lambda_0^2 + \lambda_1^2 + \lambda_2^2 + \lambda_4^2 = 1$.
The LU-invariants are
\begin{equation}
\label{4blu2}
J_1 = \lambda_1^2 \lambda_4^2 \hspace{.5cm}
J_2 = \lambda_0^2 \lambda_2^2 \hspace{.5cm}
J_4 = \lambda_0^2 \lambda_4^2.
\end{equation}
Eq.(\ref{4blu2}) implies that the Groverian measure for Eq.(\ref{4b2}) is independent of
the phase factor $\varphi$ like type 4a.

\subsection{Type4c ($\lambda_1=0$)}
In this case the state vector $|\psi \rangle$ in Eq.(\ref{state1})
reduces to
\begin{equation}
\label{4c1}
|\psi\rangle = \lambda_0 |000\rangle + \lambda_2 |101\rangle
+ \lambda_3 |110 \rangle + \lambda_4 |111\rangle
\end{equation}
with $\lambda_0^2 + \lambda_2^2 + \lambda_3^2 + \lambda_4^2 = 1$.
The LU-invariants in this type are
\begin{eqnarray}
\label{4clu1}
& &J_1 = \lambda_2^2 \lambda_3^2 \hspace{.5cm}
   J_2 = \lambda_0^2 \lambda_2^2 \hspace{.5cm}
   J_3 = \lambda_0^2 \lambda_3^2
                                             \\   \nonumber
& &J_4 = \lambda_0^2 \lambda_4^2 \hspace{.5cm}
   J_5 = 2 \lambda_0^2 \lambda_2^2 \lambda_3^2.
\end{eqnarray}
From Eq.(\ref{4clu1}) it is easy to show
\begin{equation}
\label{4clu2}
J_1 (J_2 + J_3 + J_4) + J_2 J_3 = \sqrt{J_1 J_2 J_3} = \frac{1}{2} J_5.
\end{equation}

In this type $\vec{r}_1$, $\vec{r}_2$ and $g_{ij}$ defined in Eq.(\ref{def41}) are
\begin{eqnarray}
\label{4ccom1}
& &\vec{r}_1 = (0, 0, 2 \lambda_0^2 - 1)        \\  \nonumber
& &\vec{r}_2 = (2 \lambda_2 \lambda_4, 0, \lambda_0^2 + \lambda_2^2 - \lambda_3^3 - \lambda_4^2)
                                                               \\   \nonumber
& &g_{ij} = \left(             \begin{array}{ccc}
                   2 \lambda_0 \lambda_3  &  0  &  0      \\
                   0  &  -2 \lambda_0 \lambda_3  &  0     \\
                   -2 \lambda_2 \lambda_4  &  0  &  1 - 2 \lambda_2^2
                               \end{array}                            \right).
\end{eqnarray}
Like type 4a and type 4b solving Eq.(\ref{algebraic1}) is highly
non-trivial mainly due to non-diagonalization of $g_{ij}$. Of
course, the fact that the first component of $\vec{r}_2$ is
non-zero makes hard to solve Eq.(\ref{algebraic1}) too. The
explicit expressions of the Groverian measure in this type are not
given in this paper.

\subsection{Type5 (real states): $\varphi = 0$, $\pi$}

\subsubsection{$\varphi = 0$}
In this case the state vector $|\psi \rangle$ in Eq.(\ref{state1})
reduces to
\begin{equation}
\label{501}
|\psi\rangle = \lambda_0 |000\rangle + \lambda_1 |100\rangle
+ \lambda_2 |101 \rangle + \lambda_3 |110 \rangle + \lambda_4 |111\rangle
\end{equation}
with $\lambda_0^2 + \lambda_1^2 + \lambda_2^2 + \lambda_3^2 + \lambda_4^2 = 1$.
The LU-invariants in this case are
\begin{eqnarray}
\label{500lu1}
& &J_1 = (\lambda_2 \lambda_3 - \lambda_1 \lambda_4)^2 \hspace{.5cm}
   J_2 = \lambda_0^2 \lambda_2^2 \hspace{.5cm}
   J_3 = \lambda_0^2 \lambda_3^2
                                             \\   \nonumber
& &J_4 = \lambda_0^2 \lambda_4^2 \hspace{.5cm}
   J_5 = 2 \lambda_0^2 \lambda_2 \lambda_3 (\lambda_2 \lambda_3 - \lambda_1 \lambda_4).
\end{eqnarray}
It is easy to show $\sqrt{J_1 J_2 J_3} = J_5 / 2$.

\subsubsection{$\varphi = \pi$}
In this case the state vector $|\psi \rangle$ in Eq.(\ref{state1})
reduces to
\begin{equation}
\label{502}
|\psi\rangle = \lambda_0 |000\rangle - \lambda_1 |100\rangle
+ \lambda_2 |101 \rangle + \lambda_3 |110 \rangle + \lambda_4 |111\rangle
\end{equation}
with $\lambda_0^2 + \lambda_1^2 + \lambda_2^2 + \lambda_3^2 + \lambda_4^2 = 1$.
The LU-invariants in this case are
\begin{eqnarray}
\label{500lu2}
& &J_1 = (\lambda_2 \lambda_3 + \lambda_1 \lambda_4)^2 \hspace{.5cm}
   J_2 = \lambda_0^2 \lambda_2^2 \hspace{.5cm}
   J_3 = \lambda_0^2 \lambda_3^2
                                             \\   \nonumber
& &J_4 = \lambda_0^2 \lambda_4^2 \hspace{.5cm}
   J_5 = 2 \lambda_0^2 \lambda_2 \lambda_3 (\lambda_2 \lambda_3 + \lambda_1 \lambda_4).
\end{eqnarray}
It is easy to show $\sqrt{J_1 J_2 J_3} = J_5 / 2$ in this type.

The analytic calculation of $P_{max}$ in type 5 is most difficult problem.
In addition, we don't know whether it is mathematically possible or not. However,
the geometric interpretation of $P_{max}$ presented in Ref.\cite{tama07-1} and
Ref.\cite{tama08-1} may provide us valuable insight. We hope to leave this issue
for our future research work too. The results in this section is summarized
in Table I.

\begin{center}
\begin{tabular}{c|c|c|c}  \hline
\multicolumn{2}{c|} {Type} & conditions & $P_{max}$  \\ \hline \hline
\multicolumn{2}{c|} {Type I}    &  $J_i = 0$  &  $1$       \\  \hline
{}  &   {}   & $J_i = 0$ except $J_1$  &  $\frac{1}{2} \left(1 + \sqrt{1 - 4 J_1} \right)$  \\
                                                                               \cline{3-4}
Type II  &  a  & $J_i = 0$ except $J_2$  & $\frac{1}{2} \left(1 + \sqrt{1 - 4 J_2} \right)$ \\
                                                                               \cline{3-4}
{}   &  {}  & $J_i = 0$ except $J_3$   & $\frac{1}{2} \left(1 + \sqrt{1 - 4 J_3} \right)$ \\
                                                                               \cline{2-4}
{}   &  b  &  $J_i = 0$ except $J_4$  &  $\frac{1}{2} \left(1 + \sqrt{1 - 4 J_4} \right)$ \\
                                                                               \hline
{}  &  a  &  $\lambda_1 = \lambda_4 = 0$  &  $\scriptstyle \frac{1}{4}
\left(1 + \sqrt{1 - 4(J_1 + J_2)} + \sqrt{1 - 4(J_1 + J_3)} +  \sqrt{1 - 4(J_2 + J_3)} \right)$
                                                           if $a^2 \geq b^2 + c^2$  \\
                                                                                \cline{4-4}
{}  & {}  &  {}  &  $4 \sqrt{J_1 J_2 J_3} / \left(4(J_1 + J_2 + J_3) - 1 \right)$
                                                        if $a^2 \leq b^2 + c^2$  \\
                                                                                \cline{2-4}
Type III  & {} & $\lambda_1 = \lambda_2 = 0$ & $\frac{1}{2} \left(1 +
                                                 \sqrt{1 - 4 (J_3 + J_4)} \right)$  \\
                                                                                \cline{3-4}
{} & b  &  $\lambda_1 = \lambda_3 = 0$ & $\frac{1}{2} \left(1 +
                                                 \sqrt{1 - 4 (J_2 + J_4)} \right)$  \\
                                                                                \cline{3-4}
{} & {} & $\lambda_2 = \lambda_3 = 0$ & $\frac{1}{2} \left(1 +
                                                 \sqrt{1 - 4 (J_1 + J_4)} \right)$  \\
                                                                                \hline
{} & a & $\lambda_4 = 0$ & independent of $\varphi$: not presented  \\
                                                                                \cline{2-4}
Type IV & b & $\lambda_2 = 0$ & independent of $\varphi$: not presented  \\
                                                                                \cline{3-4}
{} & {} &  $\lambda_3 = 0$  &  independent of $\varphi$: not presented  \\
                                                                                \cline{2-4}
{} & c  & $\lambda_1 = 0$  &  not presented  \\
                                                                                \hline
\multicolumn{2}{c|} {Type V} & $\varphi = 0$ & not presented  \\
                                                                                \cline{3-4}
\multicolumn{2}{c|} {} & $\varphi = \pi$ & not presented         \\
                                                                                \hline

\end{tabular}

\vspace{0.1cm}
Table I: Summary of $P_{max}$ in various types.
\end{center}
\vspace{0.5cm}

\section{New Type}

\subsection{standard form}
In this section we consider new type in $3$-qubit states. The type we consider is
\begin{equation}
\label{newstate1}
|\Phi\rangle=a|100\rangle+b|010\rangle+c|001\rangle+q|111\rangle,\quad
a^2+b^2+c^2+q^2=1.
\end{equation}
First, we would like to derive the standard form like Eq.(\ref{state1}) from
$|\Phi\rangle$.
This can be achieved as following. First, we consider LU-transformation of
$|\Phi\rangle$, {\it i.e.} $(U \otimes \openone \otimes \openone) |\Phi\rangle$,
where
\begin{eqnarray}
\label{newunitary1}
U = \frac{1}{\sqrt{a q + bc}}
\left(           \begin{array}{cc}
               \sqrt{a q} e^{i \theta}   &  \sqrt{b c} e^{i \theta}  \\
               -\sqrt{b c}               &   \sqrt{a q}
                  \end{array}                             \right).
\end{eqnarray}
After LU-transformation, we perform Schmidt decomposition following Ref.\cite{acin00}.
Finally we choose $\theta$ to make all $\lambda_i$ to be positive.
Then we can derive the standard form (\ref{state1}) from $|\Phi \rangle$ with
$\varphi = 0$ or $\pi$, and
\begin{eqnarray}
\label{relacoe1}
& &\lambda_0 = \sqrt{\frac{(ac + bq) (ab + cq)}{aq + bc}}  \\  \nonumber
& &\lambda_1 = \frac{\sqrt{abcq}}{\sqrt{(ab + cq) (ac + bq) (aq + bc)}}
               |a^2 + q^2 - b^2 - c^2|                    \\   \nonumber
& &\lambda_2 = \frac{1}{\lambda_0} |ac - bq|              \\   \nonumber
& &\lambda_3 = \frac{1}{\lambda_0} |ab - cq|              \\   \nonumber
& &\lambda_4 = \frac{2 \sqrt{abcq}}{\lambda_0}.
\end{eqnarray}
It is easy to prove that the normalization condition $a^2 + b^2 + c^2 + q^2 = 1$
guarantees the normalization
\begin{equation}
\label{newnormal1}
\lambda_0^2 + \lambda_1^2 + \lambda_2^2 + \lambda_3^2 + \lambda_4^2 = 1.
\end{equation}
Since $|\Phi\rangle$ has three free parameters, we need one more constraint between
$\lambda_i$'s. This additional constraint can be derived by trial and error. The
explicit expression for this additional relation is
\begin{equation}
\label{additional1}
\lambda_0^2 (\lambda_2^2 + \lambda_3^2 + \lambda_4^2 )
= \frac{1}{4} - \frac{\lambda_1^2}{\lambda_4^2} (\lambda_2^2 + \lambda_4^2)
(\lambda_3^2 + \lambda_4^2).
\end{equation}
Since all $\lambda_i$'s are not vanishing but there are only three free
parameters, $|\Phi \rangle$ is not involved in the types discussed in the previous
section.

\subsection{LU-invariants}
Using Eq.(\ref{relacoe1}) it is easy to derive LU-invariants which are
\begin{eqnarray}
\label{newlu-1}
& &J_1 = (\lambda_1 \lambda_4 - \lambda_2 \lambda_3)^2
                           = \frac{1}{(ab + cq)^2 (ac + bq)^2}    \\   \nonumber
& & \hspace{1.5cm}
\times
\left[ 2 abcq |a^2 + q^2 - b^2 - c^2| - (aq + bc) |(ab - cq) (ac - bq)| \right]^2
                                                                  \\  \nonumber
& & J_2 = \lambda_0^2 \lambda_2^2 = (ac - bq)^2                  \\   \nonumber
& & J_3 = \lambda_0^2 \lambda_3^2 = (ab - cq)^2                  \\   \nonumber
& & J_4 = \lambda_0^2 \lambda_4^2 = 4 abcq                       \\   \nonumber
& & J_5 = \lambda_0^2 \left( J_1 + \lambda_2^2 \lambda_3^2 - \lambda_1^2 \lambda_4^2
                                                   \right).
\end{eqnarray}
One can show directly that $J_5 = 2 \sqrt{J_1 J_2 J_3}$. Since $|\Phi \rangle$ has
three free parameters, there should exist additional relation between $J_i$'s. However,
the explicit expression may be hardly derived. In principle, this constraint can be
derived as following. First, we express the coefficients $a$, $b$, $c$, and $q$ in
terms of $J_1$, $J_2$, $J_3$ and $J_4$ using first four equations of Eq.(\ref{newlu-1}).
Then the normalization condition $a^2 + b^2 + c^2 + q^2 = 1$ gives explicit
expression of this additional constraint. Since, however, this procedure requires
the solutions of quartic equation, it seems to be hard to derive it explicitly.

Since $J_1$ contains absolute value, it is dependent on the regions in the
parameter space. Direct calculation shows that $J_1$ is
\begin{eqnarray}
\label{newj1}
J_1 = \left\{            \begin{array}{l}
              (aq - bc)^2
 \hspace{1.0cm}
           \mbox{when} \hspace{.2cm} (a^2 + q^2 - b^2 - c^2) (ab - cq) (ac - bq) \geq 0
                                                                    \\
              (aq - bc)^2 \left[1 + 2 (ab - cq) (ac - bq) (aq + bc) /
                                         (ab + cq) (ac + bq) (aq - bc)  \right]^2
                                                                 \\ \hspace{3.0cm}
           \mbox{when} \hspace{.2cm} (a^2 + q^2 - b^2 - c^2) (ab - cq) (ac - bq) < 0.
                        \end{array}                      \right.
\end{eqnarray}
Since $P_{max}$ is manifestly LU-invariant quantity, it is obvious
that it also depends on the regions on the parameter space.

\subsection{calculation of $ P_{max} $}

$P_{max}$ for state $|\Phi \rangle$ in Eq.(\ref{newstate1}) has
been analytically computed recently in Ref.\cite{tama08-1}. It
turns out that $P_{max}$ is differently expressed in three
distinct ranges of definition in parameter space. The final
expressions can be interpreted geometrically as discussed in
Ref.\cite{tama08-1}. To express $P_{max}$ explicitly we define
\begin{eqnarray}
\label{share1}
& &r_1 \equiv b^2 + c^2 - a^2 - q^2  \hspace{1.0cm}
   r_2 \equiv a^2 + c^2 - b^2 - q^2
                                                \\   \nonumber
& &r_3 \equiv a^2 + b^2 - c^2 - q^2  \hspace{1.0cm}
\omega \equiv ab + qc  \hspace{1.0cm} \mu \equiv ab - qc.
\end{eqnarray}

The first expression of $P_{max}$, which can be expressed in terms of circumradius of
convex quadrangle is
\begin{equation}
\label{share2}
P_{max}^{(Q)} = \frac{4(ab + qc) (ac + qb) (aq + bc)}{4 \omega^2 - r_3^2}.
\end{equation}
The second expression of $P_{max}$, which can be expressed in terms of circumradius of
crossed-quadrangle is
\begin{equation}
\label{shhare3}
P_{max}^{(CQ)} = \frac{(ab - cq) (ac - bq) (bc - aq)}{4 S_x^2}
\end{equation}
where
\begin{equation}
\label{share4}
S_x^2 = \frac{1}{16} (a + b + c + q) (a + b - c - q) (a - b + c - q) (-a + b + c - q).
\end{equation}
The final expression of $P_{max}$ corresponds to the largest coefficient:
\begin{equation}
\label{share5}
P_{max}^{(L)} = \mbox{max} (a^2, b^2, c^2, q^2)
              = \frac{1}{4} \left(1 + |r_1| + |r_2| + |r_3| \right).
\end{equation}
The applicable domain for each $P_{max}$ is fully discussed in Ref.\cite{tama08-1}.

Now we would like to express all expressions of $P_{max}$ in terms
of LU-invariants. For the simplicity we choose a simplified case,
that is $(a^2 + q^2 - b^2 - c^2) (ab - cq) (ac - bq) \geq 0$. Then
it is easy to derive
\begin{eqnarray}
\label{share6}
& &r_1^2 = 1 - 4(J_2 + J_3 + J_4)  \hspace{1.0cm} r_2^2 = 1 - 4(J_1 + J_3 + J_4)
                                                                 \\   \nonumber
& &r_3^2 = 1 - 4(J_1 + J_2 + J_4)  \hspace{1.0cm} \omega^2 = J_3 + J_4.
\end{eqnarray}
Then it is simple to express $P_{max}^{(Q)}$ and $P_{max}^{(CQ)}$ as following:
\begin{eqnarray}
\label{share7}
& &P_{max}^{(Q)} = \frac{4 \sqrt{(J_1 + J_4) (J_2 + J_4) (J_3 + J_4)}}
                        {4(J_1 + J_2 + J_3 + 2 J_4) - 1}
                                                          \\  \nonumber
& &P_{max}^{(CQ)} = \frac{4 \sqrt{J_1 J_2 J_3}}{4 (J_1 + J_2 + J_3 + J_4) - 1}.
\end{eqnarray}
If we take $q=0$ limit, we have $\lambda_4 = J_4 = 0$. Thus $P_{max}^{(Q)}$ and
$P_{max}^{(CQ)}$ reduce to $4 \sqrt{J_1 J_2 J_3} / (4 (J_1 + J_2 + J_3) - 1)$,
which exactly coincides with $P_{max}^<$ in Eq.(\ref{pmax53}). Finally Eq.(\ref{share6})
makes $P_{max}^{(L)}$ to be
\begin{equation}
\label{share8}
P_{max}^{(L)} = \frac{1}{4} \left( 1 + \sqrt{1 - 4(J_2 + J_3 + J_4)} +
                \sqrt{1 - 4(J_1 + J_3 + J_4)} +
                              \sqrt{1 - 4(J_1 + J_2 + J_4)} \right).
\end{equation}
One can show that $P_{max}^{(L)}$ equals to $P_{max}^{>}$ in Eq.(\ref{pmax51}) when
$q=0$. This indicates that our results (\ref{share7}) and (\ref{share8}) have correct
limits to other types of three-qubit system.

\section{conclusion}
We tried to compute the Groverian measure analytically in the various types of
three-qubit system. The types we considered in this paper are given in Ref.\cite{acin00}
for the classification of the three-qubit system.

For type 1, type 2 and type 3 the Groverian measures are
analytically computed. All results, furthermore, can be
represented in terms of LU-invariant quantities. This reflects the
manifest LU-invariance of the Groverian measure.

For type 4 and type 5 we could not derive the analytical
expressions of the measures because the Lagrange multiplier
equations (\ref{algebraic1}) is highly difficult to solve.
However, the consideration of LU-invariants indicates that the
Groverian measure in type 4 should be independent of the phase
factor $\varphi$. We expect that this fact may drastically
simplify the calculational procedure for obtaining the analytical
results of the measure in type 4. The derivation in type 5 is most
difficult problem. However, it might be possible to get valuable
insight from the geometric interpretation of $P_{max}$, presented
in Ref.\cite{tama07-1} and Ref.\cite{tama08-1}. We would like to
revisit type 4 and type 5 in the near future.

We think that the most important problem in the research of entanglement is to understand
the general properties of entanglement measures in arbitrary qubit systems. In order to
explore this issue we would like to extend, as a next step, our calculation to
four-qubit states. In addition, the Groverian measure for four-qubit pure state is
related to that for two-qubit mixed state via purification\cite{shapira06}. Although
general theory for entanglement is far from complete understanding at present stage,
we would like to go toward this direction in the future.

{\bf Acknowledgement}:
This work was supported by the Kyungnam University
Research Fund, 2007.

\newpage

\begin{appendix}{\centerline{\bf Appendix A}}

\setcounter{equation}{0}
\renewcommand{\theequation}{A.\arabic{equation}}

One can easily show that the elements of ${\cal O}$ defined in Eq.(\ref{lu1})
are given by
\begin{eqnarray}
\label{app1}
& &{\cal O}_{11} = \frac{1}{2} \left( u_{11} u_{22}^* + u_{11}^* u_{22} +
               u_{12} u_{21}^* + u_{12}^* u_{21} \right)
                                                \\   \nonumber
& &{\cal O}_{22} = \frac{1}{2} \left( u_{11} u_{22}^* + u_{11}^* u_{22} -
               u_{12} u_{21}^* - u_{12}^* u_{21} \right)
                                                \\   \nonumber
& &{\cal O}_{33} = |u_{11}|^2 - |u_{12}|^2    \\   \nonumber
& &{\cal O}_{12} = \frac{i}{2} \left( u_{12} u_{21}^* + u_{11} u_{22}^* -
               u_{12}^* u_{21} - u_{11}^* u_{22} \right)
                                                \\   \nonumber
& &{\cal O}_{21} = \frac{i}{2} \left( u_{12} u_{21}^* + u_{11}^* u_{22} -
               u_{12}^* u_{21} - u_{11} u_{22}^* \right)
                                                \\   \nonumber
& &{\cal O}_{13} = u_{11} u_{12}^* + u_{11}^* u_{12}
                                                     \\   \nonumber
& &{\cal O}_{31} = u_{11} u_{21}^* + u_{11}^* u_{21}
                                                      \\   \nonumber
& &{\cal O}_{23} = -i \left( u_{11} u_{12}^* + u_{21}^* u_{22} \right)
                                                    \\   \nonumber
& &{\cal O}_{32} = i \left( u_{11} u_{21}^* + u_{12}^* u_{22} \right)
\end{eqnarray}
where $u_{ij}$ is element of the unitary matrix defined in Eq.(\ref{lu1}). It is
easy to prove ${\cal O} {\cal O}^T = {\cal O}^T {\cal O} = \openone$, which
indicates that ${\cal O}_{\alpha \beta}$ is an element of O(3).
\end{appendix}

\newpage

\begin{appendix}{\centerline{\bf Appendix B}}

\setcounter{equation}{0}
\renewcommand{\theequation}{B.\arabic{equation}}
If the density matrix associated from the pure state $|\psi \rangle$ in
Eq.(\ref{state1}) is represented by Bloch form like Eq.(\ref{density1}), the
explicit expressions for $\vec{v}_i$ are
\begin{eqnarray}
\label{appb1}
& &\vec{v}_1 = \left(     \begin{array}{c}
                         2 \lambda_0 \lambda_1 \cos \varphi  \\
                         2 \lambda_0 \lambda_1 \sin \varphi  \\
         \lambda_0^2 - \lambda_1^2 - \lambda_2^2 - \lambda_3^2 - \lambda_4^2
                          \end{array}
                                                    \right)
\hspace{1.0cm}
\vec{v}_2 = \left(     \begin{array}{c}
            2 \lambda_1 \lambda_3 \cos \varphi + 2 \lambda_2 \lambda_4 \\
                        - 2 \lambda_1 \lambda_3 \sin \varphi  \\
         \lambda_0^2 + \lambda_1^2 + \lambda_2^2 - \lambda_3^2 - \lambda_4^2
                          \end{array}
                                                    \right)
                                                               \\  \nonumber
& &\hspace{3.0cm}
\vec{v}_3 = \left(     \begin{array}{c}
            2 \lambda_1 \lambda_2 \cos \varphi + 2 \lambda_3 \lambda_4 \\
                        - 2 \lambda_1 \lambda_2 \sin \varphi  \\
         \lambda_0^2 + \lambda_1^2 - \lambda_2^2 + \lambda_3^2 - \lambda_4^2
                          \end{array}
                                                    \right)
\end{eqnarray}
and the components of $h^{(i)}$ are
\begin{eqnarray}
\label{appb2}
& &h^{(1)}_{11} = 2 \lambda_2 \lambda_3 + 2 \lambda_1 \lambda_4 \cos \varphi,
\hspace{1.0cm}
h^{(1)}_{22} = 2 \lambda_2 \lambda_3 - 2 \lambda_1 \lambda_4 \cos \varphi
                                                                   \\  \nonumber
& &h^{(1)}_{33} = \lambda_0^2 + \lambda_1^2 - \lambda_2^2 - \lambda_3^2
                  + \lambda_4^2,
\hspace{1.0cm}
h^{(1)}_{12} = h^{(1)}_{21} = -2 \lambda_1 \lambda_4 \sin \varphi
                                                                   \\  \nonumber
& &h^{(1)}_{13} = -2 \lambda_2 \lambda_4 + 2 \lambda_1 \lambda_3 \cos \varphi,
\hspace{1.0cm}
h^{(1)}_{31} = -2 \lambda_3 \lambda_4 + 2 \lambda_1 \lambda_2 \cos \varphi
                                                                    \\  \nonumber
& &h^{(1)}_{23} = -2 \lambda_1 \lambda_3 \sin \varphi,
\hspace{1.0cm}
h^{(1)}_{32} = -2 \lambda_1 \lambda_2 \sin \varphi
                                                                    \\  \nonumber
& &h^{(2)}_{11} = - h^{(2)}_{22} = 2 \lambda_0 \lambda_2,
\hspace{1.0cm}
h^{(2)}_{33} = \lambda_0^2 - \lambda_1^2 + \lambda_2^2 - \lambda_3^2
                  + \lambda_4^2
                                                                     \\  \nonumber
& &h^{(2)}_{12} = h^{(2)}_{21} = 0,
\hspace{1.0cm}
h^{(2)}_{13} = 2 \lambda_0 \lambda_1 \cos \varphi
                                                                      \\  \nonumber
& &h^{(2)}_{31} = -2 \lambda_3 \lambda_4 - 2 \lambda_1 \lambda_2 \cos \varphi,
\hspace{1.0cm}
h^{(2)}_{23} = 2 \lambda_0 \lambda_1 \sin \varphi
                                                                   \\  \nonumber
& &h^{(2)}_{32} = 2 \lambda_1 \lambda_2 \sin \varphi.
\end{eqnarray}
The matrix $h^{(3)}_{\alpha \beta}$ is obtained from $h^{(2)}_{\alpha \beta}$
by exchanging $\lambda_2$ with $\lambda_3$. The non-vanishing components of
$g_{\alpha \beta \gamma}$ are
\begin{eqnarray}
\label{appb3}
& &g_{111} = -g_{122} = -g_{212} = -g_{221} = 2 \lambda_0 \lambda_4
                                                         \\  \nonumber
& &g_{113} = -g_{223} = 2 \lambda_0 \lambda_3,
\hspace{1.0cm}
g_{131} = -g_{232} = 2 \lambda_0 \lambda_2
                                                         \\  \nonumber
& &g_{133} = 2 \lambda_0 \lambda_1 \cos \varphi,
\hspace{1.0cm}
g_{233} = 2 \lambda_0 \lambda_1 \sin \varphi
                                                         \\  \nonumber
& &g_{312} = g_{321} = 2 \lambda_1 \lambda_4 \sin \varphi,
\hspace{1.0cm}
g_{311} = -2 \lambda_2 \lambda_3 - 2 \lambda_1 \lambda_4 \cos \varphi
                                                         \\   \nonumber
& &g_{313} = 2 \lambda_2 \lambda_4 - 2 \lambda_1 \lambda_3 \cos \varphi,
\hspace{1.0cm}
g_{322} = -2 \lambda_2 \lambda_3 + 2 \lambda_1 \lambda_4 \cos \varphi
                                                           \\  \nonumber
& &g_{323} = 2 \lambda_1 \lambda_3 \sin \varphi,
\hspace{1.0cm}
g_{331} = 2 \lambda_3 \lambda_4 - 2 \lambda_1 \lambda_2 \cos \varphi
                                                               \\  \nonumber
& &g_{332} = 2 \lambda_1 \lambda_2 \sin \varphi,
\hspace{1.0cm}
g_{333} = \lambda_0^2 - \lambda_1^2 + \lambda_2^2 + \lambda_3^2 - \lambda_4^2.
\end{eqnarray}

\end{appendix}
\end{document}